\documentclass[aps,prb,twocolumn,amsmath,amssymb,showpacs]{revtex4}
\usepackage{graphicx}
\usepackage{subfigure}
\usepackage{color}
\usepackage{bm}

\begin{document}
\date{\today}
\title{Stability and chaos of a driven nano-electromechanical Josephson junction}
\author{P. Berggren}
\email{Peter.Berggren@physics.uu.se}
\author{J. Fransson}
\affiliation{Department of Physics and Astronomy, Uppsala University, Box 530, SE-751 21\ \ Uppsala}

\begin{abstract}
We consider the motion of and Josephson current through a mechanically oscillating superconducting island asymmetrically embedded in a Josephson junction. The electromechanical coupling is provided by distance dependent tunneling rates between the electrodes and the island. The system asymmetry, resulting from the geometrical configuration, leads, for weak coupling, to an equation of the mechanical motion that reduces to the well-known Duffing equation. At zero bias voltage the island motion is determined by the homogenous Duffing equation that opens up two separate regions of solutions depending on the superconducting phases. The island either moves under influence of an anharmonic single well potential, or is governed by a double well potential that allows for off-center oscillations. Under applied bias voltage the island equation of motion turns into a modified Duffing equation, with time dependent coefficients, that demonstrate both quasi periodic and chaotic behavior.  
\end{abstract}
\maketitle

\section{Introduction}\label{sec:intro}
Nano electromechanical system (NEMS) resonators may now be micro fabricated precise enough that the effects of tunneling electron coupling to the mechanical system are measurable\cite{Blencowe}. Such dynamical interactions between the charge carriers and vibrational modes in a mesoscopic system have been observed in single electron tunneling to suspended carbon nano tubes\cite{Steele28082009, Lassagne28082009}. Similar effects of vibron--electron coupling possibly explain differential conductance dips and peaks in molecular electronics devices\cite{andres1996, reed1997, kergueris1999, hong2000, rosink2000, chen2000, porath2000, smit2002, reichert2002, aviram1998, langlais1999, park2000, park2002, zhitenev2002} and differential conductance steps in STM based inelastic tunneling spectroscopy on local vibration modes on surfaces\cite{stipe1998, franssonIETS2007}.

Nano scale resonator setups are interesting as fast high sensitivity detection devices\cite{knobel1} for mass\cite{DucDai2009, Naik2009, Ekinci2004, Yang2006, Peng2006, Li2007}, charge\cite{Cleland1997}, force\cite{Rugar2004} and displacement\cite{LaHaye2004} and as mechanical systems reach the quantum limit implications for quantum information technology may be tremendous\cite{cleland1, Rabl2010, Schwab2005, Blencowe2004}. 

The field has further evolved to include and explore superconducting NEMS. One investigative direction has been to couple nano mechanical resonators to a superconducting Cooper pair box\cite{LaHaye2009}, or a superconducting quantum interference device (SQUID)\cite{Etaki2008}, in order to probe and control superconducting qubits, as well as detect displacements near the quantum limit. Another course has been to study Josephson currents coupled to mechanical, or molecular, oscillators situated within the tunneling junction\cite{Zazunov2006, Gorelik2001, Zhu2006, fransson2008, fransson2010}. The oscillator then acts to shuttle Cooper pairs at resonant levels. 

In this paper we introduce a double Josephson junction that is asymmetric with respect to the oscillatory motion of a superconducting island. The dynamics of the system is captured as the mechanical motion is coupled to the electron tunneling. Apart from reproducing the expected equation of motion terms found in Ref. \onlinecite{fransson2008} the asymmetry adds a nonlinear cubic term. At zero bias voltage the island motion is consequently controlled by the relative superconducting phases through the Duffing equation. This equation, thoroughly studied in mathematics\cite{Lai2006, Marinca2010, Parlitz1985, Sheu2007}, has received a lot of attention in NEMS research since nonlinear restoring forces act on small scale resonators, see Ref. \onlinecite{Lifshitz2009} for a review. The Duffing equation also shows up in driven macro scale resonators that are geometrically similar to our setup\cite{Heagy1991, Berger1997}.   

At finite bias voltages the equation of motion is modified to include harmonically time dependent coefficients to both the linear and cubic term as well as a harmonic driving force. No studies have been published on this Duffing equation variant to our knowledge. At Josephson frequencies above and below the eigen frequency of the oscillating island regular and stable quasi periodic motion is found, wheres more resonant frequencies yield chaotic solutions. Chaos is an inherent property of the driven Duffing equation\cite{Holmes1976}.   

The importance of nonlinearities and the Duffing equation in a NEMS aspect comes from a number of suggested and investigated applications. Weak signal amplification with low noise levels based on system sensitivity near bifurcation points is one active subject\cite{Yurke1995, Siddiqi2004, Vijay2009, Karabalin2011}. Other novel experiments utilize buckled nano resonator beams that oscillate within the confines of a double well potential, typical to the Duffing equation, either to produce mechanical quantized qubit states in the resonator by cooling\cite{Savel'ev2007} or to construct mechanical memory bits that work under room temperature by controlling transitions between the potential wells\cite{Bagheri2011}.

This paper is outlined as follows. In Sec. \ref{sec:theor} a detailed description of the mechanical system is given. Here we also derive the Josephson tunneling currents dependent on island position as well as the equation of motion for the island as it couples to the tunneling Cooper pairs. In Sec. \ref{sec:res} the solutions to the island equation of motion in absence of bias voltage are presented together with numerical results for the tunneling currents under different conditions including zero bias voltage with varying superconducting phases and finite bias voltage. In Sec. \ref{sec:sum} we summarize our findings. 
\section{Theory}\label{sec:theor}
\subsection{Description of model}\label{sec:desc}
\begin{figure}[htb]
\centering
\includegraphics[width=7cm]{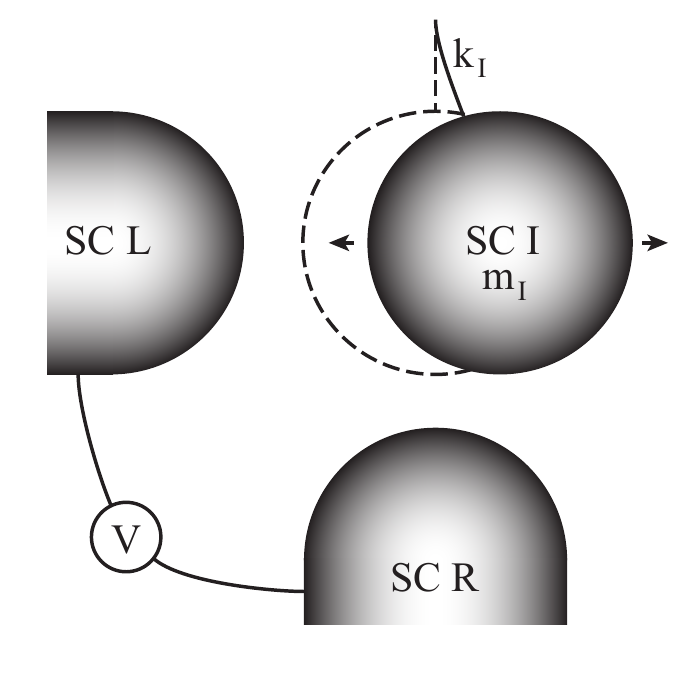}
\caption{Schematic picture of the mechanically and electronically coupled system. The superconducting island (SC I) is free to move as indicated by the arrows, while the left (SC L) and right (SC R) superconductors are rigid. $m_I$ and $k_I$ denote the island mass and spring constant respectively. A bias voltage V can be applied.}
\label{fig1}
\end{figure}

The nanomechanical system considered comprise three superconducting electrodes out of which two are fixed perpendicular to each other. At the intersection where fixed left $(L)$ and right $(R)$ electrodes point a third movable island is suspended by a cantilever. In absence of electromechanical coupling the island is allowed to vibrate in the direction of the left lead with a restoring force proportional to the distance from equilibrium. The setup is inherently asymmetric with respect to the motion of the island and an illustration of the system is shown in Fig. \ref{fig1}.

The system forms a double Josephson junction and we assume the island displacement $u$ to be small compared to the distance between the superconducting parts. Our aim is to describe the tunneling current as the electronic process couples to the mechanical motion of the island and we begin by addressing the electron Hamiltonian of the system,
\begin{equation}
H=H_L+H_R+H_I+H_T,
\label{eq:hamiltonian}
\end{equation}
where,
\begin{align}
H_{L,R,I}&=\sum_{\boldsymbol{\kappa}\sigma}\epsilon_{\boldsymbol{\kappa}}c^{\dagger}_{{\boldsymbol{\kappa}}\sigma} c_{{\boldsymbol{\kappa}}\sigma}+\sum_{\boldsymbol{\kappa}}\left[\Delta_{L,R,I}c^{\dagger}_{{\boldsymbol{\kappa}}\uparrow}c^{\dagger}_{-{\boldsymbol{\kappa}}\downarrow}+H.c.\right]
\label{eq-bcs}
\\
H_T&=\sum_{\mathbf{p}\mathbf{k}\sigma}T_{\mathbf{p}\mathbf{k}}c^{\dagger}_{\mathbf{p}\sigma}c_{\mathbf{k}\sigma}+\sum_{\mathbf{q}\mathbf{k}\sigma}T_{\mathbf{q}\mathbf{k}}c^{\dagger}_{\mathbf{q}\sigma}c_{\mathbf{k}\sigma}+H.c.
\end{align}
The first three terms of $H$ separately give the electronic structure of the leads in terms of BCS Hamiltonians, c.f. Eq. (\ref{eq-bcs}). The superconductors couple through tunneling term $H_T$. Here, $c_{\boldsymbol{\kappa}}$ and $c^\dagger_{\boldsymbol{\kappa}}$ annihilate and create an electron in electrode $\chi=L,I,R$ with momentum $\boldsymbol{\kappa}$ and spin $\sigma$. Electrons in the left, island, and right leads are denoted by momentum ${\bf p}$, ${\bf k}$, and ${\bf q}$, respectively. $\Delta_\chi$ is the superconducting pairing potential in lead $\chi$.

The tunneling response to island vibrations is modeled by distance dependent tunneling matrix elements $T_{\mathbf{p}\mathbf{k}}$ and $T_{\mathbf{q}\mathbf{k}}$. For small vibrations we use the linear approximation
\begin{equation}
T_{\mathbf{p}\mathbf{k}}=T^{(0)}_{\mathbf{p}\mathbf{k}}(1-\alpha u),
\label{eq:tleft}
\end{equation}
where $\alpha$ is a positive coupling constant and $T^{(0)}$ is the tunneling rate to the island at its equilibrium position. The matrix element for tunneling between the island and the right lead is given by 
\begin{equation}
T_{\mathbf{q}\mathbf{k}}=T^{(0)}_{\mathbf{q}\mathbf{k}}\left(1-\alpha\left[\sqrt{R^2+u^2}-R\right]\right),
\label{eq:tright}
\end{equation}
where $R$ is the equilibrium distance from the island to the right lead. 

By assuming low temperatures, $T\sim 0.01-1\mbox{K}$, we can work with vibrational energies of the uncoupled island in the range $\omega_0\sim 10^{-6} - 10^{-3}$ eV, which is small in relation to the typical electron energy of $1$ eV.
\subsection{Josephson current modulated by the island oscillation}\label{sec:current}
We derive the tunneling current, defined by $I_\chi(t)=-e\langle\dot{N}_\chi(t)\rangle$, where $N_\chi$ is the number operator, at junction $\chi(=L,R)$ to the island by following Ref. \onlinecite{fransson2010} and obtain,
\begin{equation}
\begin{split}
I_\chi(t)=&2e\mbox{Re}\int^t_{-\infty}e^{-i\omega_\chi(t+t')}\langle\left[A(t),A(t')\right]\rangle\\
&+e^{-i\omega_\chi(t-t')}\langle\left[A(t),A^{\dagger}(t')\right]\rangle dt',
\end{split}
\label{eq:current}
\end{equation}
where $\omega_\chi=\mu_\chi-\mu_I$ defines the voltage drop between lead $\chi$ and the island ($\mu_\chi$ and $\mu_I$ are the chemical potentials of lead $\chi$ and the island, respectively). The operators $A(t)=\sum_{\boldsymbol{\kappa}\mathbf{k}\sigma}T_{\boldsymbol{\kappa}\mathbf{k}}c^{\dagger}_{\boldsymbol{\kappa}\sigma}(t)c_{\mathbf{k}\sigma}(t)$,
for $\boldsymbol{\kappa}\in L,R$ and the time-dependence is defined by
\begin{equation}
\begin{split}
c_{{\boldsymbol{\kappa}}\sigma}(t)&=e^{iK_\chi t}c_{\boldsymbol{\kappa}\sigma}e^{-iK_\chi t}\\
c_{\mathbf{k}\sigma}(t)&=e^{iK_It}c_{\mathbf{k}\sigma}e^{-iK_It},
\end{split}
\end{equation}
where $K_\chi=H_\chi-\mu_\chi N_\chi$ and $K_I=H_I-\mu_IN_I$.

In Eq. (\ref{eq:current}) the junction current is divided into two terms which describe different tunneling mechanisms. The second term accounts for the single electron tunneling and will not be addressed further in this text. Our focus is here devoted to the first term, which describes the Josephson tunneling current.

We make use of the Bogoliubov--Valatin transformation $c_{\boldsymbol{\kappa}\sigma}=u_{\boldsymbol{\kappa}}\gamma_{\boldsymbol{\kappa}\sigma}-\eta\nu^*_{\boldsymbol{\kappa}}\gamma^{\dagger}_{\boldsymbol{\kappa}\bar{\sigma}}$, where $\eta=\pm1$ differs in sign for spin up or spin down electrons, whereas $u_{\boldsymbol{\kappa}}$ and $\nu_{\boldsymbol{\kappa}}$ are the coherence factors satisfying $|u_{\boldsymbol{\kappa}}|^2+|\nu_{\boldsymbol{\kappa}}|^2=1$ and $u^*_{\boldsymbol{\kappa}}\nu_{\boldsymbol{\kappa}}=|\Delta_{\chi}|e^{i\phi_{\chi}}/(2E_{\boldsymbol{\kappa}})$, where $\phi_{\chi}$ is the superconducting phase in lead $\chi$. Through the transformation, we define the quasi-particle energies
\begin{equation}
E_{\boldsymbol{\kappa}}=\sqrt{(\epsilon_{\boldsymbol{\kappa}}-\mu_{\chi})^2+|\Delta_{\chi}|^2}.
\end{equation} 
We can, thus, write
\begin{equation}
\begin{split}
\langle\left[A(t),A(t')\right]\rangle&=\sum_{\boldsymbol{\kappa}\mathbf{k}\sigma}\frac{|\Delta_{\chi}||\Delta_I|}{4E_{\boldsymbol{\kappa}}E_{\mathbf{k}}}T_{\boldsymbol{\kappa}\mathbf{k}}(t)T_{\boldsymbol{\kappa}\mathbf{k}}(t')\\
&\times\left(e^{i(E_{\boldsymbol{\kappa}}+E_{\mathbf{k}})\tau}-e^{-i(E_{\boldsymbol{\kappa}}+E_{\mathbf{k}})\tau}\right)e^{-i\phi_{\chi}},
\end{split}
\label{eq:avbog}
\end{equation}
where $\tau=t-t'$ and $\phi_{\chi}=\phi_I-\phi_{\chi}$. Our assumptions of small vibrational energies justifies the approximation $T_{\mathbf{p}\mathbf{k}}(t')=T_{\mathbf{p}\mathbf{k}}(t)-\tau\dot{T}_{\mathbf{p}\mathbf{k}}(t)$, which leads to the Josephson current $I_L(t)$ from the left lead to the island
\begin{equation}
\begin{split}
I_L(t)=&J_L[1-\alpha u]^2\sin{(\omega_{J,L}t+\phi_L)}\\
&+\Gamma_L[1-\alpha u]\alpha\dot{u}\cos{(\omega_{J,L}t+\phi_L)}.
\end{split}
\label{eq:currentleft}
\end{equation}
Here, the amplitudes
\begin{subequations}
\begin{align}
J_{\chi}(eV)=&e\sum_{\boldsymbol{\kappa}\mathbf{k}}|T^{(0)}_{\boldsymbol{\kappa}\mathbf{k}}|^2\frac{|\Delta_{\chi}||\Delta_I|}{2E_{\boldsymbol{\kappa}}E_{\mathbf{k}}}
\nonumber\\
&\times\left(\frac{1}{eV+E_{\boldsymbol{\kappa}}+E_{\mathbf{k}}}-\frac{1}{eV-E_{\boldsymbol{\kappa}}-E_{\mathbf{k}}}\right),
\\
\Gamma_{\chi}(eV)=&e\sum_{\boldsymbol{\kappa}\mathbf{k}}|T^{(0)}_{\boldsymbol{\kappa}\mathbf{k}}|^2\frac{|\Delta_{\chi}||\Delta_I|}{2E_{\boldsymbol{\kappa}}E_{\mathbf{k}}}
\nonumber\\
&\times\left(\frac{1}{(eV+E_{\boldsymbol{\kappa}}+E_{\mathbf{k}})^2}-\frac{1}{(eV-E_{\boldsymbol{\kappa}}-E_{\mathbf{k}})^2}\right),
\end{align}
\end{subequations}
define the tunneling between the fixed electrode $\chi=L,R$ and the island in absence and presence of the coupling to the vibrational mode, respectively.

The tunneling from the right lead to the island is given by Eq. (\ref{eq:currentleft}) after replacing the tunneling matrix element Eq. (\ref{eq:tleft}) with Eq. (\ref{eq:tright}), i.e.
\begin{equation}
\begin{split}
I_R(t)=&J_R\left(1-\alpha\left[\sqrt{R^2+u^2}-R\right]\right)^2\sin{(\omega_{J,R}t+\phi_R)}\\
&+\Gamma_R\left(1-\alpha\left[\sqrt{R^2+u^2}-R\right]\right)\\
&\times\left(\frac{\alpha\dot{u}u}{\sqrt{R^2+u^2}}\right)\cos{(\omega_{J,R}t+\phi_R)}.
\end{split}
\label{eq:currentright}
\end{equation}

\subsection{Island motion modulated by the electron coupling}\label{sec:islandmotion}
In addition to the cantilever spring force acting on the island, electromechanical coupling contributes with a dynamical force. To model this more complicated equation of motion we construct a Hamiltonian that include energy terms $H_{J,L}$ and $H_{J,R}$ originating from coupling in each junction,
\begin{equation}
H_{osc}=H^{(0)}_{osc}+H_{J,L}+H_{J,R},
\end{equation}
where $H^{(0)}_{osc}=\frac{p^2}{2m_I}+\frac{k_Iu^2}{2}$. $p$ is the island momentum, while $m_I$ denotes its mass, and $k_I$ is the cantilever spring constant.   

$H_{J,L}$ and $H_{J,R}$ are constructed  out of the requirement,
\begin{equation}
2e\frac{\partial H_J}{\partial\phi}=I_J,
\end{equation}     
which is fulfilled if,
\begin{equation}
\begin{split}
H_{J,L}=&\frac{J_L}{2e}[1-\alpha u]^2\left(1-\cos{(\omega_{J,L}t+\phi_L)}\right)\\
&-\frac{\Gamma_L}{2em_I}[1-\alpha u]\alpha p\sin{(\omega_{J,L}t+\phi_R)}
\end{split}
\end{equation}
and
\begin{equation}
\begin{split}
H_{J,R}=&\frac{J_R}{2e}\left(1-\alpha\left[\sqrt{R^2+u^2}-R\right]\right)^2\\
&\times\left(1-\cos(\omega_{J,R}t+\phi_R)\right)\\
&+\frac{\Gamma_R}{2em_I}\left(1-\alpha\left[\sqrt{R^2+u^2}-R\right]\right)\\
&\times\left(\frac{\alpha pu}{\sqrt{R^2+u^2}}\right)\sin(\omega_{J,R}t+\phi_R).
\end{split}
\end{equation}

With a complete Hamiltonian the full island motion, $u$, is obtained by solving the Hamilton equations of motion,
\begin{equation}
\dot{u}=\frac{\partial H_{osc}}{\partial p},\qquad\dot{p}=-\frac{\partial H_{osc}}{\partial u}.
\label{eq:hamilton}
\end{equation}
In doing so we arrive at the following differential equation,
\begin{equation}
m_I\ddot{u}+(\gamma_L+\gamma_{R,2})\dot{u}+(k_I+\gamma_{R,1})u=F_L,
\label{eq:ofmotion}
\end{equation}   
where,
\begin{equation}
\begin{split}
\gamma_L=&-\frac{\Gamma_L\alpha^2}{e}\sin{(\omega_{J,L}t+\phi_L)},\\
\gamma_{R,2}=&\frac{\Gamma_R\alpha}{e}\left[(1+\alpha R)\left(\frac{1}{\sqrt{R^2+u^2}}-\frac{u^2}{[R^2+u^2]^{3/2}}\right)-\alpha\right]\\
&\times\sin{(\omega_{J,R}t+\phi_R)},\\
\gamma_{R,1}=&\frac{J_R\alpha}{e}\left(\frac{1+\alpha R}{\sqrt{R^2+u^2}}-\alpha\right)\\
&\times\left[\left(\frac{\Gamma_R\omega_{J,R}}{2J_R}+1\right)\cos{(\omega_{J,R}t+\phi_R)}-1\right],\\
\intertext{and}
F_L=&\frac{-J_L\alpha}{e}(1+\alpha u)\left[\left(\frac{\Gamma_L\omega_{J,L}}{2J_L}+1\right)\cos{(\omega_{J,L}t+\phi_L)}-1\right].
\end{split}
\label{eq:faktorer}
\end{equation}
One may note that equation (\ref{eq:ofmotion}) lacks a driving force term, $F_R$, that will be present if the angle between the right lead and the island motion differs from $90^{\circ}$. 

The central island equation of motion contains both time and nontrivial position dependence in its coefficients. A more transparent equation is found in the weak coupling and low bias voltage limit. Under such conditions $\alpha$ is small and $\Gamma_{\chi}\ll J_{\chi}$, so terms proportional to either $\Gamma_{\chi}$, $\alpha\Gamma_{\chi}$ or $\alpha^2$ are dropped to enlighten the terms of greatest physical relevance. 

We also bear in mind that $u/R\ll1$ and keep only the second order Taylor expansions, 
\begin{equation}
\frac{1}{\sqrt{R^2+u^2}}\simeq\frac{1}{R}-\frac{u^2}{2R^3}
\end{equation}
etc.

The coefficients (\ref{eq:faktorer}) approximate to,
\begin{equation}
\begin{split}
\gamma_L&\approx0,\\
\gamma_{R,2}&\approx0,\\
\gamma_{R,1}&\approx\frac{J_R\alpha}{eR}\left(\cos{(\omega_{J,R}t+\phi_R)}-1\right)\left(1-\frac{u^2}{2R^2}\right),\\
F_L(t)&\approx-\frac{J_L\alpha}{e}\left(\cos{(\omega_{J,L}t+\phi_L)}-1\right),
\end{split}
\label{eq:faktorersimp}
\end{equation}
and in defining,
\begin{equation}
\begin{split}
A(t)=&\frac{1}{m_I}[k_I+k_D(\cos(\omega_{J,R}t+\phi_R)-1)]\\
B(t)=&\frac{k_D}{2m_IR^2}(\cos(\omega_{J,R}t+\phi_R)-1)
\end{split}
\end{equation}
where, $k_D=(J_R\alpha)/(eR)$, act as a dynamical spring constant, we end up at the equation of motion,
\begin{equation}
\ddot{u}+A(t)u-B(t)u^3=F_L(t)/m_I.
\label{eq:ofmotionsimp}
\end{equation} 
This is a Duffing equation modified by time dependent coefficients. It is only analytically solvable for zero bias voltage, through series expansions\cite{Pelster2003}, or by Jacobi's elliptic functions\cite{Duffingsol}.   
\section{Results and discussion}\label{sec:res}
The dynamics of equation (\ref{eq:ofmotionsimp}) directly effects the overall Josephson current through its solutions. Due to the equations nonlinear nature we approach these numerically in the general case and analytically for zero bias voltage.  
\subsection{System under zero bias voltage}
\begin{figure}[t]
\centering
\includegraphics[width=8cm]{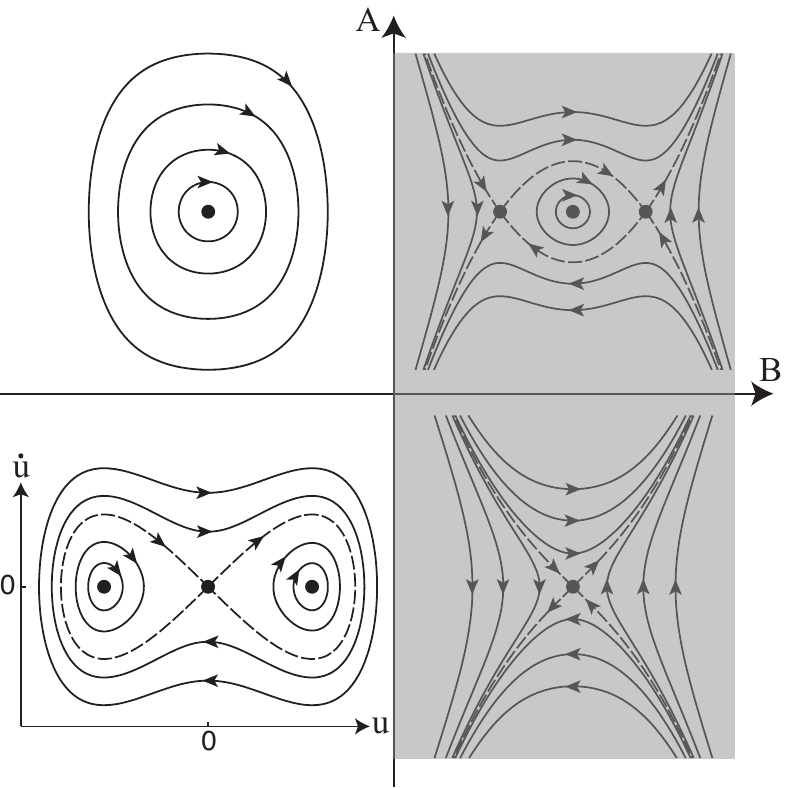}
\caption{Zero bias voltage phase diagrams of four characteristic solution regions to the equation of motion (\ref{eq:ofmotionsimp}) as it depends on the coefficients $A$ and $B$ when $F=0$. $B$ is always negative and only solutions from leftmost quadrants are physical.}
\label{fig2}
\end{figure}

Even at zero bias voltage, $A(t),B(t),F_L(t)\rightarrow A,B,F$, the island equation of motion has a rich variety of solutions if the superconducting phase differences are nonzero. Depending on the coefficients sign in equation (\ref{eq:ofmotionsimp}) this Josephson effect opens up two distinct regions of solutions, depicted in the phase diagrams of Fig. \ref{fig2}. While four regions of solutions are obtainable mathematically, only two are physical since $B$ is non-positive.

The force term $F$ is non-negative and acts to shift the island motion away from the left lead, consequently lowering the DC tunneling rate. The following analytic solutions apply to the $\phi_L=0$ case where $F=0$. Note that the current $I_L=0$ at all times under such conditions within the approximations made above.  

Also note that $\Gamma_L$ and $\Gamma_R$ are zero when no bias voltage is applied which means that the second term in the current expressions vanish. 

In terms of energy the island is confined by the potential $V(u)=\frac{A}{2}u^2-\frac{B}{4}u^4$ which for $A>0$, $B<0$ is a single well. Under these conditions the equation of motion (\ref{eq:ofmotionsimp}) has one singular point of center type and all phase trajectories are closed. The equation is satisfied by the solution,
\begin{equation}
u=u_0\mbox{cn}(\Omega t,k), 
\end{equation}
where $u_0$ is the amplitude, $\mbox{cn}(x,y)$ is the Jacobi elliptic cosine, $\Omega=\sqrt{A-Bu_0^2}$, and $k=\sqrt{-B/2}\cdot u_0/\Omega$. See upper left quadrant of Fig. \ref{fig2}.

If $A<0$, $B<0$ the island motion is governed by a double well potential which give rise to three singular points of which two are centers, corresponding to the double well bottoms, while the third is of saddle type centered between the two wells. All phase trajectories are closed and as is clear from Fig. \ref{fig2} solutions exist that circumfere either one of the two singular points of center type as well as solutions that enclose all three singular points. For $A<0,B<0$ solutions to equation (\ref{eq:ofmotionsimp}) can be written,
\begin{equation}
u=\left\{
\begin{array}{ll}
\pm u_0\mbox{dn}(\omega_1t,k_1)&\mbox{for }\sqrt{\frac{|A|}{(-B)}}<u_0<\sqrt{\frac{2|A|}{(-B)}}\\[1em]
u_0\mbox{cn}(\omega_2t,k_2)&\mbox{for }u_0>\sqrt{\frac{2|A|}{(-B)}},
\end{array}
\right.
\end{equation}    
where $\omega_1=\sqrt{-B/2}u_0$, $\omega_2=\sqrt{-Bu_0^2-|A|}$, $k_1=\omega_2/\omega_1$, $k_2=\omega^2_1/\omega^2_2$ and dn is a Jacobi elliptic function. 

\begin{figure}[b]
\includegraphics[width=7cm]{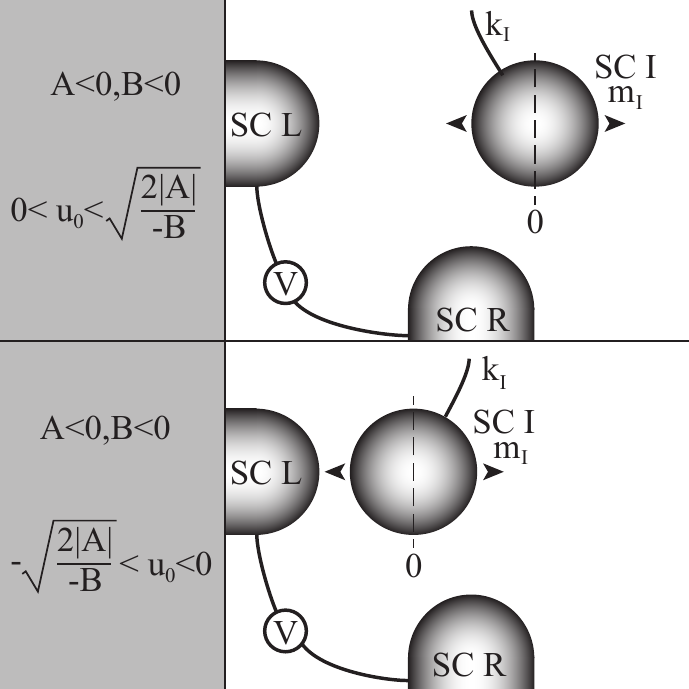}
\caption{Schematic image of the island trapped in one of two potential wells present in the $A<0,B<0$ case, at zero bias voltage and small phase shifts $\phi_L$. For the indicated limits on $u_0$ the island oscillates with its center either to the right (above) or to the left (below).}
\label{fig4}
\end{figure}

The upper solutions above correspond to trajectories enclosing either one of the two singular points of center type. A sign change on the initial condition $u_0$, within the limit, gives rise to oscillations of equal frequency whose origin is separated by a distance $2\sqrt{|A|/(-B)}$ in real space. A schematic picture of the two solutions are depicted in Fig. \ref{fig4}.
The different solutions will not change the tunneling current between the island and right lead but the DC component of the tunneling current between the left lead and the island is clearly affected. At $\phi_L=0$, or in other words $F_L=0$, this effect is absent since equation (\ref{eq:currentleft}) yields zero current. For a small phase shift $\phi_L\neq 0$, on the other hand, both solutions exist together with a non zero tunneling current $I_L$. The magnitude of the DC tunneling current difference between solutions confined to the two separate potential wells is less than $\Delta I_{L,DC}<J_L4\alpha\sqrt{|A|/(-B)}\sin{\phi_L}$.

\begin{figure}[t]
\centering
\includegraphics[width=8cm]{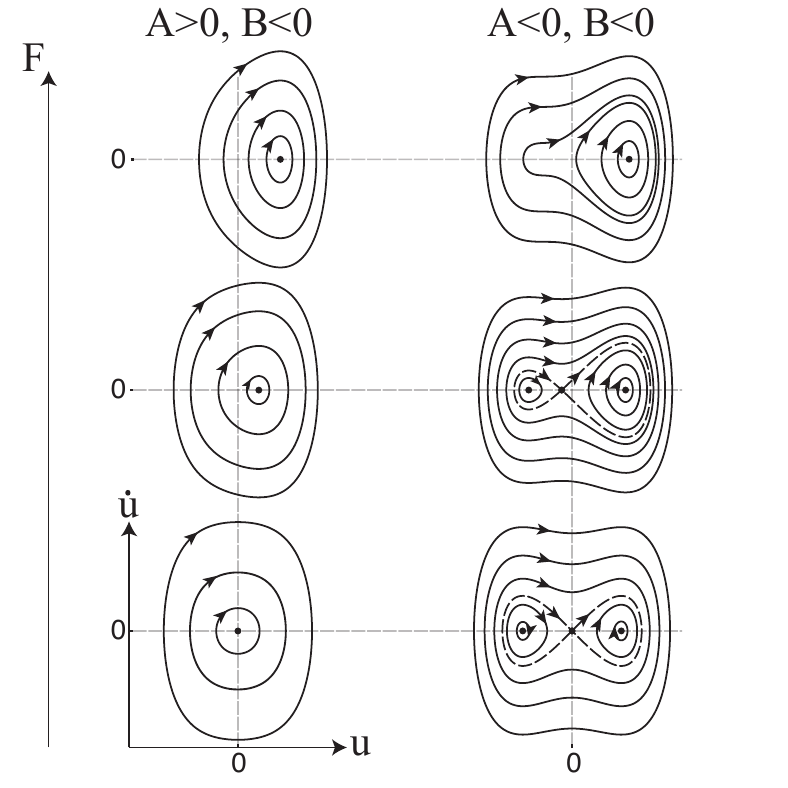}
\caption{Phase portraits of the two possible solution regions as they depend on the size of the force term, $F$. The centre in the $A>0,B<0$ case shifts toward the right while the phase trajectories become more elliptical with a major axis parallel to the velocity. In the $A<0,B<0$ case the leftmost two centers vanish with growing $F$ while the rightmost center slowly shift towards the right.}
\label{fig3}
\end{figure}

Figure \ref{fig3} illustrates how phase shifts $\phi_L$ distorts solution trajectories as well as moves the singular points. On the negative side of the origin the singular point of center type moves in positive direction, as $F$ grows larger, while the singular point of saddle type moves in negative direction. The two points eventually merge, leaving only the singular point of centre type on the positive side. This point, on the other hand, slowly shifts to more positive values as $\phi_L$ increases.

For $A>0$, $B<0$ the singular point of center type shifts towards positive values as $\phi_L$ increases at the same time as the phase trajectories distorts toward an elliptical shape with major axis along the $\dot{u}$ direction.

One may also note that while the island oscillations pass the origin the frequency of the motion induced $I_R$ is double that of $I_L$ due to the system geometry. As soon as oscillations are restricted to either the positive or negative side the tunneling currents have equal frequency.  

To analyze the Josephson tunneling under coupling to the mechanical motion of the island, at zero bias voltage, the Fourier transform is taken for a fine mesh of varying phase shifts $\phi_R$ and $\phi_L$. For $\phi_L=0$ the analytical solutions are used while a numerical solver is utilized when $\phi_L\neq0$. As far as values goes the results are to be taken qualitatively even though realistic input parameters are used. 

The current parameters and the coupling constant are set to\cite{Zhu} $J_L=J_R=0.1\hspace{1mm}\mbox{mA}$ and $\alpha=0.01\hspace{1mm}\mbox{\AA}^{-1}$. As equilibrium distance between the leads $R=10\hspace{1mm}\mbox{\AA}$ is taken, while an island mass of $m_I=1\hspace{1mm}\mbox{fg}$ is used.

\begin{figure}[t]
\centering
\includegraphics[width=7cm]{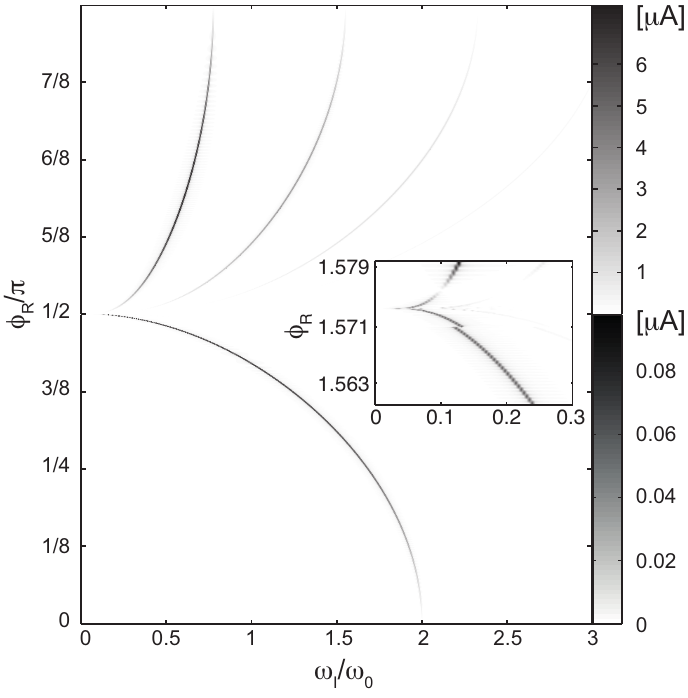}
\caption{Fourier transform of the Josephson tunneling current $I_R$ as a function of its frequency $\omega_I$ and the phase shift $\phi_R$. $\omega_0=6.6\cdot10^{-8}\hspace{1mm}\mbox{eV}$ is the island eigenfrequency and $\phi_L=0$ at all times. The figure is composed of two images divided as the color bars indicate. Both color bars are graded in $\mu A$. The lower half correspond to single well solutions where $A>0$ and $B<0$ while the upper half stems from solutions confined to one of the the double wells in the $A<0$ and $B<0$ case. The inset is an enlargement of a small area around $\phi_R/\pi=1/2$.} 
\label{fig5}
\end{figure}

Figure \ref{fig5} depicts the Fourier transform of the tunneling current $I_R$ as a function of its frequency and the phase shift $\phi_R$ for a given initial value $u_0=0.1\hspace{1mm}\mbox{nm}$ and mechanical spring constant $k_{I}=0.01\hspace{1mm}\mbox{N/m}$. Unless stated otherwise these are the input values used in calculations throughout the remainder of this paper. As a result the energy associated with the eigenfrequency of the uncoupled island is $\omega_0=6.6\cdot10^{-8}\hspace{1mm}\mbox{eV}$. All figures of Fourier transformed currents are given without DC component. As $\phi_R$ increases from 0 to $\pi/2$ solutions to the island equation of motion are restricted to the $A>0$, $B<0$ region, while phase values between $\phi_R=\pi/2$ and $\pi$ result in solutions within the $A<0$, $B<0$ region having trajectories that enclose the positive singular point of center type. Tunneling currents are close to singly harmonic in the $0<\phi_R<\pi/2$ region and with increasing phase shift the frequency drops until it approaches zero as $A$ vanishes when $\phi_R=\pi/2$. The value $\phi_R=\pi/2$ has no significance in it self but stems from the parameter input $k_I=k_D$. 

Above $\phi_R=\pi/2$ the island motion has a more complicated shape which is reflected in the larger number of harmonics needed for its description. The current amplitude, given in $\mu A$, is noticeably two orders of magnitude larger compared to the amplitude of the bottom half arc. This is expected for solutions to the island motion with trajectories enclosing one of the singular points of center type when the initial value lies close to the origin for the $A<0,B<0$ case in comparison to the $A>0,B<0$ case. The AC current lies superimposed on top of the DC current and while solution trajectories depicted in the bottom left quadrant of Fig. \ref{fig2} moves away from the origin all the way around the singular point to the right, trajectories in the upper left quadrant never go beyond the initial value.   

The inset of Fig. \ref{fig5} depicts a small area around $\phi_r=\pi/2$ and indicates that the transition between the two arc structures is not direct. As $A$ goes from positive to negative the island motion phase portrait in the lower left quadrant of Fig. \ref{fig2} builds up from the origin. The two singular points of center type divide from the the single center and leaves a saddle point behind. This can be seen as a discontinues step in the lower arc of the inset and solution trajectories now enclose all three singular points. With bigger $\phi_R$ values $A$ becomes more negative and the lying eight shape of the phase portrait grows. Eventually the separatrix curve reaches the initial value $u_0$ where the island nears the origin infinitesimally slow and the frequency drops to zero. Shortly after this point is passed the oscillation amplitude drops to zero as the initial value $u_0$ and the bottom of the double well potential coincide.

\begin{figure}[b]
\centering
\includegraphics[width=7cm]{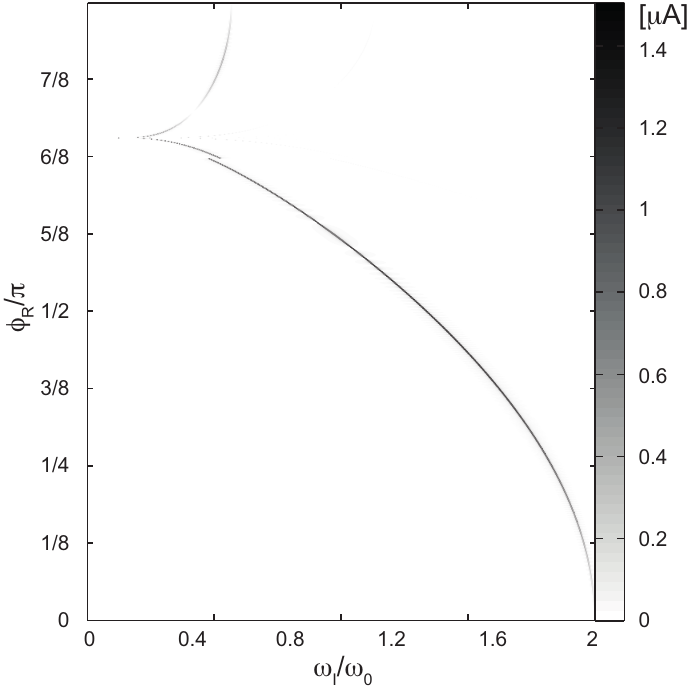}
\caption{Fourier transform of $I_R$ as in Fig. \ref{fig5} with initial condition and spring constant changed to $u_0=0.4\hspace{1mm}\mbox{nm}$ and $k_I=0.017\hspace{1mm}\mbox{N/m}$, which leads to $\omega_0=8.6\cdot10^{-8}\hspace{1mm}\mbox{eV}$. Currents from the additional set of island motion solutions, enclosing all three singular points in the $A<0,B<0$ region, clearly show up above the discontinues step in the lower arc.} 
\label{fig6}
\end{figure}

The differences in amplitude between solution regions even out when input values are changed to $u_0=0.4\hspace{1mm}\mbox{nm}$ and $k_I=0.017\hspace{1mm}\mbox{N/m}$, in accord with the discussion above. For these values the transitions of the Fig. \ref{fig5} inset happens over a larger phase shift $\phi_R$ range as well. All portrayed in Fig. \ref{fig6}.

\begin{figure*}[t]
\centering
\includegraphics[width=\textwidth]{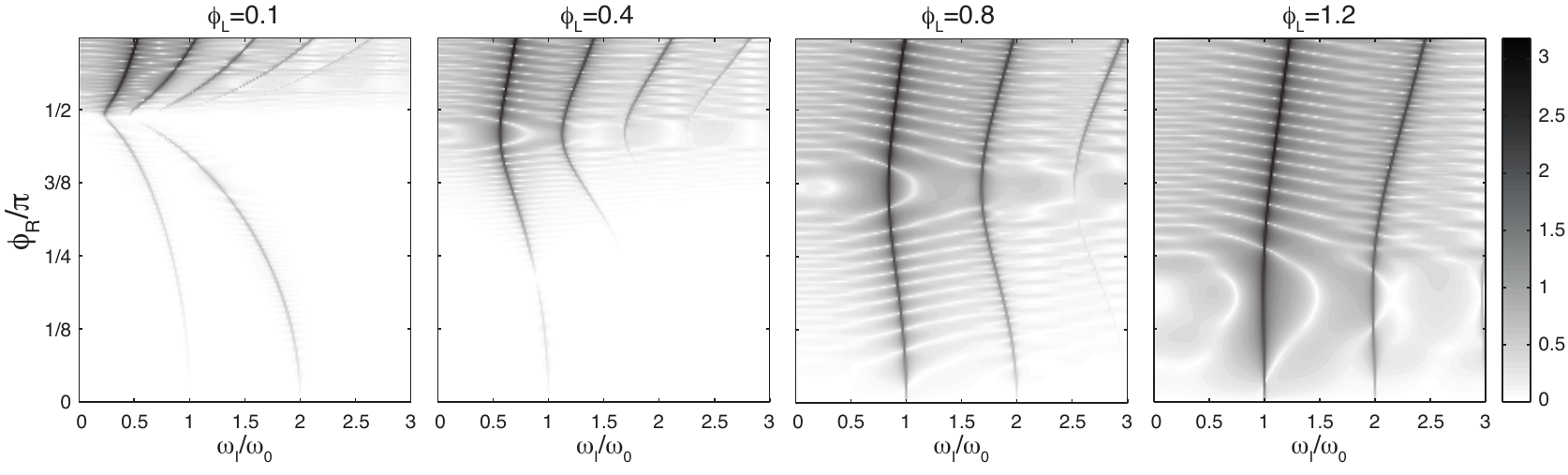}
\caption{The Fourier transform of the Josephson tunneling current $I_R$ as a function of its frequency $\omega_I$ and phase shift $\phi_R$ depicted with increasing phase shifts $\phi_L$ from left to right. The amplitude scale is logarithmic on the form $\mbox{log}_{10}(1+I_R*10^{8})$.}
\label{fig8}
\end{figure*}

With non zero phase shifts, $\phi_L\neq0$, the force term $F$ in equation (\ref{eq:ofmotionsimp}) becomes finite positive which changes the island vibration signature in the tunneling current. Figure \ref{fig8} illustrates this with four consecutive images where $\phi_L$ increases for each image from left to right.

First off the current amplitude from solutions in the $A>0,B<0$ region varies with growing $\phi_L$ and secondly the distinctive features of each solution region become less pronounced. The changes can be understood with figure \ref{fig3} in mind. In the first two images of Fig. \ref{fig8} the stable center in the $A>0,B<0$ region moves in the positive direction toward the initial value $u_0$ and the amplitude diminishes. In the following two images the stable center has passed $u_0$ and amplitude gets bigger. A nonzero $\phi_L$ complicates the picture further since it causes $\phi_R$ to also shift the stable center towards more positive values. This is evident in the $\phi_L=0.4$ image of Fig. \ref{fig8} where the stable center passes the initial value $u_0$ just below $\phi_R=\pi/4$. 

The tunneling current frequency never goes to zero in either image of Fig. \ref{fig8} even though the double well potential governs the island motion shortly above $\phi_R=\pi/2$, and upwards, in the $\phi_L=0.1$ case. The positive force term $F$ shifts the $\phi_R$ value at which separation of the singular points occur. When this happens the center point of the single well potential has already passed the initial value $u_0$ and the separatrix curve is never crossed. In the remaining three images with higher $\phi_L$ values the double well potential never form and no separatrix curve appears.    

All images in Fig. \ref{fig8} seem to indicate that the tunneling current frequencies perfectly match at the transition from region $A>0,B<0$ to $A<0,B<0$ in contrast to the $\phi_L=0$ case. The discontinuity in Fig. \ref{fig5} is very small on the other hand and numerical noise makes it hard to distinguish such fine features. \subsection{System under finite bias voltage}
A bias voltage $\omega_{J,L}=-\omega_{J,R}>0$ over the junction setup significantly changes the character of the tunneling current. Most noticeably the island motion is no longer strictly periodic, but rather quasi periodic or even chaotic. The current amplitude also becomes some three orders of magnitude greater than in the zero bias case. This is easily understood as the usual Josephson factor in expressions (\ref{eq:currentleft}) and (\ref{eq:currentright}) varies between $-1$ and $1$ while the factor associated with island motion changes in the order of $10^{-3}$. All figures presented below are obtained with $\phi_l=\phi_R=0$.  

At low bias voltage, $\omega_J/\omega_0<0.24$ $(\omega_J<1.6\cdot10^{-8}\hspace{1mm}\mbox{eV})$, the island motion follows a regular pattern where it is quasi periodic with a high frequency, low amplitude, oscillation superimposed on a higher amplitude vibration of frequency equal to $\omega_J$. This causes small ripples on, as well as distorts, the tunneling current dominated by the Josephson factor. The typical case situation is shown in the bottom four images of Fig. \ref{fig9}.

The current induced coefficients in the island equation of motion are zero at 
\begin{equation}
t=0+n\cdot2\pi/\omega_J\hspace{10mm}n=0,1,2,\dots
\end{equation} 
where the island vibration is harmonic. The cantilever spring dictates the motion for a few periods until the nonlinear and driving force contributions rapidly grow in sync with the Josephson AC. In such time intervals of strong nonlinearity the phase portrait implies a stable center markedly shifted away from the left SC lead.  

Between $0.24<\omega_J/\omega_0<6.1$ $(1.6\cdot10^{-8}\hspace{1mm}\mbox{eV}<\omega_J<4.0\cdot10^{-7}\hspace{1mm}\mbox{eV})$ the harmonic driving force and island motion frequencies are comparable, but instead of simply resonating, the island vibrates in a chaotic fashion. This behavior comes as no surprise for such a strongly nonlinear driven system. Chaotic solutions are a well studied property of the ordinary driven Duffing equation but here we only go as far as to compare Poincar\'{e} maps taken at $\omega_J/\omega_0=0.48$ and $\omega_J/\omega_0=24$ to conclude that the low voltage map is compliant with a characteristic chaotic map while the higher voltage map has a clearly quasi periodic structure. The middle four images of Fig. \ref{fig9} illustrates the island motion and Josephson tunneling current  in the chaotic region at $\omega_J/\omega_0=0.48$ $(\omega_J=3.2\cdot10^{-8}\hspace{1mm}\mbox{eV})$.
\begin{figure}[h]
\centering
\includegraphics[width=8 cm]{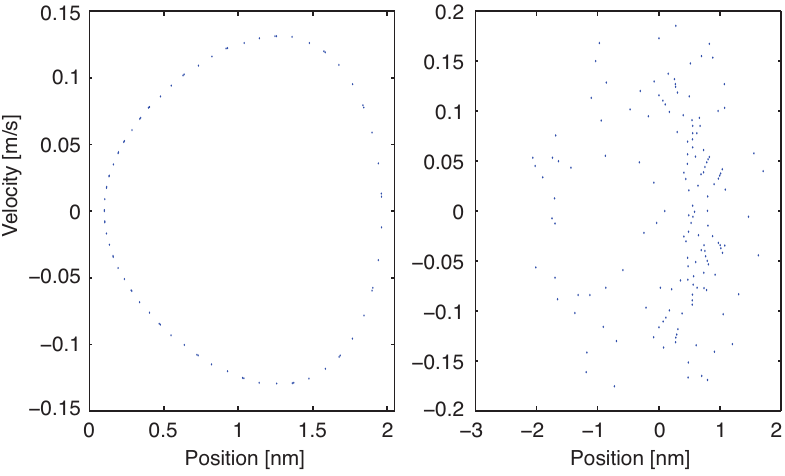}
\caption{Poincar\'{e} maps of the island motion taken at the $t = 0+n\cdot2\pi/\omega_J\hspace{5 mm}n=0,1,2,\dots$ intersection and bias voltages $\omega_J/\omega_0=24$ (left) and $\omega_J/\omega_0=0.48$ (right).}
\label{fig9}
\end{figure}

Above $\omega_J/\omega_0>6.1$ $(\omega_J>4.0\cdot10^{-7}\hspace{1mm}\mbox{eV})$ regularity in the island motion reappears, as the top four images of Fig. \ref{fig9}, taken at $\omega_J/\omega_0=24$ $(\omega_J=1.6\cdot10^{-6}\hspace{1mm}\mbox{eV})$, indicate. In this region the Josephson frequency is higher than that of the major oscillatory island motion and the tunneling current is subject to a slow modulation.      	

In contrast to the low bias voltage case, where the momentary phase portraits vary adiabatically with respect to the island vibrations, high bias voltage causes rapid changes in the time dependent equation of motion coefficients. The comparatively slow island is subject to a quickly deforming single well whose bottom shifts from the origin to a finite positive value with period $T=2\pi/\omega_J$. With increasing bias voltages the fine wave pattern in the solution trajectories seen in the top left image of Fig. \ref{fig9} diminishes. 
\begin{figure*}[t]
\centering
\includegraphics[width=\textwidth]{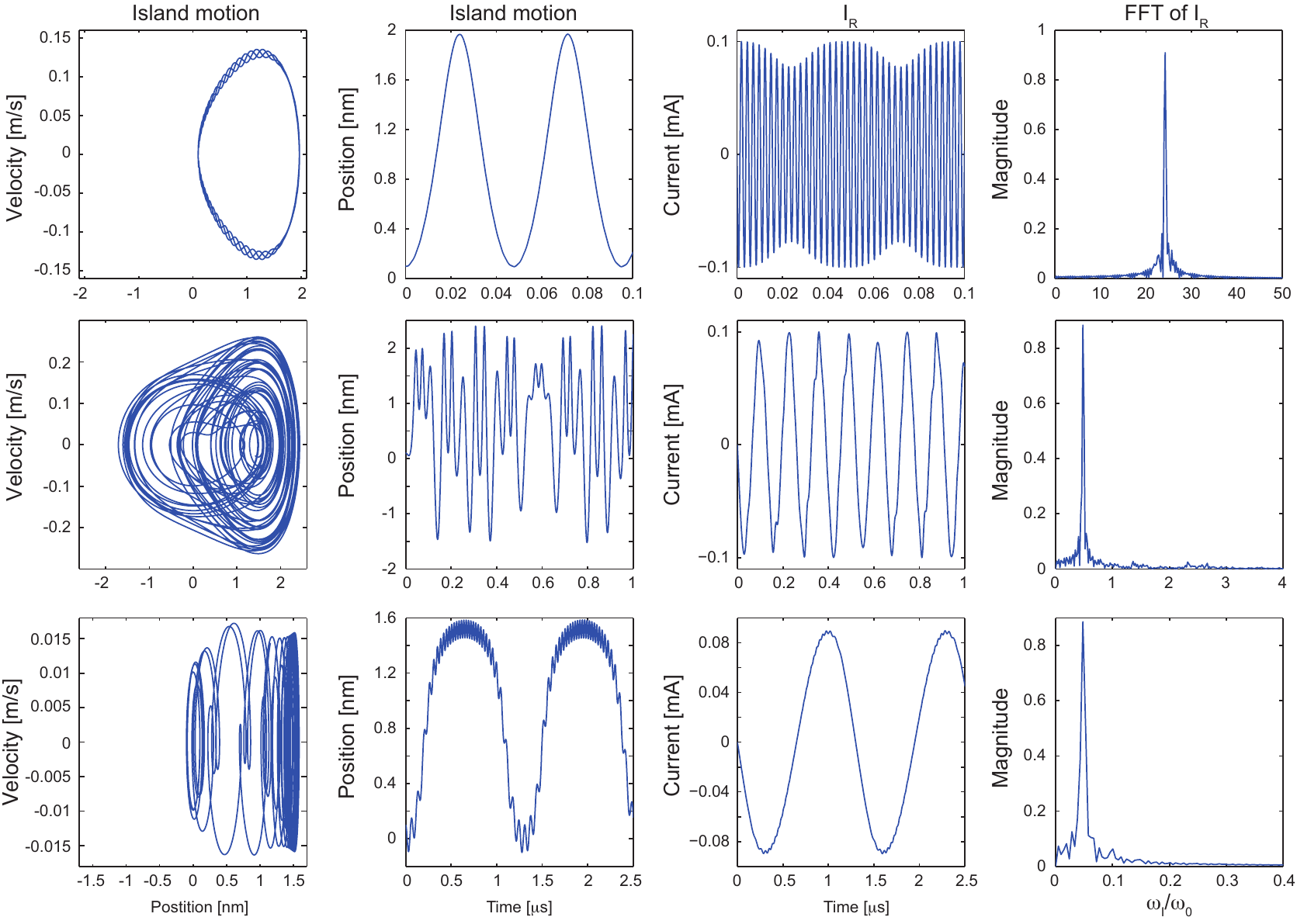}
\caption{Images depicting the phase portrait of the island motion, the island position as a function of time, the Josephson tunneling current as a function of time, and the fast Fourier transform of the current from left to right at bias voltages $\omega_J/\omega_0=24$ (top), $\omega_J/\omega_0=0.48$ (middle), and $\omega_J/\omega_0=0.048$ (bottom).}
\label{fig9}
\end{figure*}
\section{Summary}\label{sec:sum}
We have studied electron tunneling coupled to a mechanical oscillator in a double Josephson junction. The geometry of the setup is asymmetric with respect to the mechanical motion which introduces a nonlinear term in the oscillator equation of motion. 

The Josephson tunneling current over one junction is modeled as linearly dependent on the oscillator displacement directly while the gap width of the second junction changes as a hypothenuse to the displacement. An immediate consequence is that the mechanically induced current frequency over the first junction is half that of the second junction if the oscillator passes its equilibrium position and that the frequencies are equal if the vibrations are restricted to either the positive or negative side. 

In the uncoupled system the oscillator is taken to vibrate harmonically with a linear restoring force. Coupling adds a linear, cubic and force term to the equation of motion - all time dependent at finite bias voltage. In the zero bias voltage limit the differential equation reduces to the Duffing equation if the super conducting phases differ. 

The homogenous Duffing equation has two sets of physical solution regions in our setup. One with a single well potential if the linear term is positive and one with a double well potential if the linear term is negative. Which of the double wells the oscillator is vibrating in is indistinguishable by looking at the alternating tunneling current. The DC contributions will however vary between the two. 

Superconducting phase shifts associated with the junction in line with the direction of oscillations govern the force term and can be manipulated to shift the potential well bottoms away from the rigid SC lead. The phase shift associated with the second junction controls the linear and cubic term and a sweep through 0 to $\pi$ reveal that single well solutions are obtained at low phase shifts while larger values produce a double well potential if the mechanical spring constant is chosen properly. 

At nonzero bias voltage we find three domains of solutions to the island equation of motion with their own characteristics. For very low bias voltage, such that the Josephson frequency is low compared to the vibration frequency, the island motion is quasi periodic which distorts the tunneling current by superimposing small ripples on the current. In the intermediate voltage span, where the Josephson and island frequencies are of the same order, the system turns chaotic and the tunneling current gets irregular distortions. For larger voltages, such that the Josephson frequency is much larger than the vibration frequency, the island motion is again quasi-periodic. The current is roughly harmonic, however, with a slow modulation superimposed arising from the mechanical motion of the island.

The present study is based on theoretical assumption that should be within the realms of the state-of-the-art experimental capabilities. Detection of chaotic dynamics in nanoscale systems would be interesting from many perspectives and it is with great confidence we anticipate experimental verification of our proposal.    \section{Acknowledgements}\label{sec:ack}
We like to acknowledge the support of the Swedish Research Council.

\end{document}